\documentstyle[12pt]{article}
\title{Equation of Motion of Small Bodies in Relativity}
\author{J\"urgen Ehlers\thanks{E-mail: ehlers@aei-potsdam.mpg.de}
\\Max-Planck-Institut fuer Gravitationsphysik
\\Albert-Einstein-Institut, Am Muehlenberg 1
\\14476 Golm, Germany
\\Robert Geroch\thanks{E-mail: geroch@midway.uchicago.edu}
\\Enrico Fermi Institute, 5640 Ellis Ave,
\\Chicago, IL, 60637, USA}

\begin{document}
\maketitle

\noindent{\bf Abstract}
\\
\noindent {\sl There is proven a theorem, to the effect that a material body
in general relativity, in a certain limit of sufficiently small
size and mass, moves along a geodesic.}    
\\

Within the general theory of relativity, matter is typically
described in the following manner.
The local state of the matter, at each
event of space-time, is characterized by the values there
of certain fields on space-time; while
the dynamical evolution of that matter is then given by a
certain system of partial differential equations on those
fields.  Now consider a body composed of such matter.  
Since the system of differential equations on the matter fields
normally manifests an initial-value formulation, it follows that every
detail of the future behavior of that body is determined, say, from
given initial data for that system.

Fortunately, such a detailed description of the behavior 
of a material body is in many cases not needed:
What is often more interesting
is the ``behavior of the body as a whole".  Thus, for
the case of the Earth, we might
be interested only in the Earth's overall orbit about the Sun, 
and not in the tides, continental drift, etc.  In order to
obtain such a description, it is generally necessary to consider a
limit of a family of actual material bodies, where that 
limit involves small overall size of the body, small overall
mass of the body, and, possibly, restrictions on the mass-density 
or other properties of the body.  We expect that, on passing to a 
suitable limit along these lines, the motion of such a body
as a whole will be described by a timelike geodesic in space-time.  
Indeed, the idea that, in some suitable limit, material  
bodies move on geodesics was an important
one already at the beginnings of relativity \cite{AE}.  

It is notoriously difficult even to state a conjecture, in
general relativity, reflecting these ideas.   For example,
neither the ``total mass" nor the ``size" of a material
body arise as natural concepts within the theory, making it awkward 
to formulate the necessary limitations on the structure of the
body.  Even more difficult is the problem of how to characterize
the motion of the body as a whole.  What curve, within the world-tube
of the body, will be chosen as ``representative" of the body's
overall motion?  Will the geodesic character of that curve
be with respect to the actual metric of the space-time (which
includes, of course, the effects of the body itself), or
with respect to some ``background metric"?  If the latter,
how is this background to be defined?  Or, alternatively, should
the conjecture refer, not to some specific curve, but rather to
the ``average behavior" of the body? 

Despite these difficulties, there have been obtained, in general 
relativity, a number of results to the effect that material
bodies, in a suitable limit, must move on geodesics.
For a discussion of various results  and approaches
to this problem, see, e.g., \cite{JE}; and for a summary of
later developments see \cite{TD}, \cite{LB}.

In one result in particular, \cite{GJ}, there is derived geodesic 
behavior with respect to a background metric for a body whose
gravitational field is ignored, i.e., under the assumption
that the background metric remains fixed during passage to
the limit of a small body.  We shall, essentially, generalize
this result to the case in which the body is permitted to
manifest a (suitably small) gravitational field of its own,
in accordance with the idea outlined above.
In more detail, we shall prove the following.
\\

\noindent {\bf Theorem.}  {\em Let $M$ be a 4-manifold, $g_{ab}$ a 
smooth Lorentz-signature metric on $M$, and $\gamma$ a smooth 
timelike curve on $M$.  Consider a 
closed neighborhood $U$ of $\gamma$ and any 
neighborhood $\hat{U}$ of $g_{ab}$ in $C^1[U]$.  Let
there exist, for every such $U$, if sufficiently small,
and every such $\hat{U}$, 
a Lorentz-signature metric $\tilde{g}_{ab} \in \hat{U}$ 
whose Einstein tensor  
i) satisfies the dominant energy condition everywhere in $U$,
ii) is nonzero in some neighborhood of $\gamma$, and iii) vanishes
on $\partial U$.
Then $\gamma$ is a $g$-geodesic.}
\\

Think of $U$ as a ``world-tube" surrounding $\gamma$.  A
neighborhood, $\hat{U}$, of $g_{ab}$, in the space $C^1[U]$,
may be described as 
follows.  Fix, at each point $p$ of $U$, a neighborhood of
$g_{ab}|_p$ (in the space of symmetric tensors at $p$), and
a neighborhood of $\nabla_a|_p$ (in the space of derivative
operators at $p$), where these neighborhoods vary continuously,
but otherwise arbitrarily, from point to point in $U$.  Then the metric 
$\tilde{g}_{ab}$, for membership in this neighborhood
$\hat{U}$, must be ``close to $g_{ab}$", 
in the following sense:
At each point of $U$, the value of $\tilde{g}_{ab}$ must lie within 
the given neighborhood of $g_{ab}$ there, and derivative operator,
$\tilde{\nabla}_a$, of $\tilde{g}_{ab}$ must lie within
the given neighborhood of $\nabla_a$ there.  Thus, e.g.,
it follows that, for sufficiently
small $\hat{U}$, the curve $\gamma$ will again be timelike with respect
to every metric $\tilde{g}_{ab}$ in $\hat{U}$.  Note that the
metric $\tilde{g}_{ab}$ is defined only within the neighborhood
$U$ of the curve $\gamma$, and not outside; and that we 
restrict only the value and first derivative of $\tilde{g}_{ab}$,
and not any higher derivatives.  On the
Einstein tensor, $\tilde{G}_{ab}$, of the metric $\tilde{g}_{ab}$,
we impose, in condition i), the following energy condition:  
$\tilde{G}_{ab}\tilde{t}^a\tilde{t}'^b \geq 0$ for any
two future-directed $\tilde{g}$-timelike vectors, $\tilde{t}^a$ and 
$\tilde{t}'^b$.  This implies that, for fixed $\tilde{t}^a$ and
$\tilde{t}'^b$, {\em every} component of $\tilde{G}_{ab}$ is bounded
by a suitable multiple of $\tilde{G}_{ab}\tilde{t}^a\tilde{t}'^b$,
where that multiple depends, of course, on the frame with respect
to which the components of $\tilde{G}_{ab}$ are taken.    

Think of the metric $\tilde{g}_{ab}$ as a solution of Einstein's
equation representing a massive body (condition ii)) confined
to a neighborhood of $\gamma$ (condition iii)).  The theorem
contemplates the existence of a sequence of such solutions, which
approach the given ``background" metric $g_{ab}$, in this
$C^1$-sense.  Note that we do {\em not} require that the 
stress-energies of the $\tilde{g}_{ab}$ approach the stress-energy
of $g_{ab}$ (for that would require $C^2$-convergence).  
Indeed, the stress-energies of the $\tilde{g}_{ab}$
could be unbounded during the approach to $g_{ab}$.  [An
example of this behavior is that with $M, g_{ab}$ Minkowski space-time,
$\gamma$ a timelike geodesic therein, and the $\tilde{g}_{ab}$
the metrics of Schwarzschild fluid balls (of successively
smaller radii $R$), centered on $\gamma$
and with mass given by $m \propto R^{5/2}$.]  Note also that
we do not make any assumptions about the stress-energy of 
$g_{ab}$ itself.  In any
case, the theorem asserts that, under these conditions, $\gamma$    
is a geodesic with respect to $g_{ab}$.  Note that this implies,
in particular, that the $\tilde{g}$-accelerations of $\gamma$
approach zero.  It is in this sense, then, that the theorem 
asserts that ``small massive bodies move on near-geodesics". 
\\

\noindent {\bf Proof of the Theorem:}  Denote by $u^a$ the unit 
tangent to $\gamma$, and set 
$A^a = u^m\nabla_m u^a$, the acceleration of $\gamma$.
Let, for contradiction, $p_0$ be a point of $\gamma$ at
which $A^a \neq 0$.  

Choose vector fields $t^a$, $x^a$, and $\beta^a$, defined
in a neighborhood of $p_0$, such that:  i) at $p_0$, $t^a = u^a$,
$x^a = A^a/|A^b|$ and
$\beta^a = 0$; ii) $t^a$, $x^a$  and $\beta^a$ are transported
along $\gamma$ according to the laws $u^m\nabla_m t^a = 0$,
$u^m\nabla_m x^a = 0$, and
$u^m\nabla_m\beta^a = 2 u_b\ x^{[b}t^{a]}$, respectively; and iii) each
of these three vector fields has, everywhere on $\gamma$, 
vanishing symmetrized derivative.
[Thus, each of $t^a$, $x^a$, and $\beta^a$
is ``Killing on $\gamma$".  Near the point $p_0$, $t^a$ 
and $x^a$ behave like ``translations"; while $\beta^a$ behaves
like a ``boost".]  Choose
points $p_+$ and $p_-$ of $\gamma$, lying on either side
of point $p_0$, such that  
\begin{equation}
\beta^a = f_+ t^a + g x^a,
\label{alph+}\end{equation}\begin{equation}  
\beta^a = f_- t^a - g x^a,
\label{alph-}\end{equation}
at $p_+$ and $p_-$, respectively, where $f_+ >0$, $f_- >0$ 
and $g$ are numbers.  [That such points exist follows from 
conditions i) and ii), above, on 
$t^a$, $x^a$, and $\beta^a$.  Indeed, it follows immediately 
from these two conditions that, everywhere on $\gamma$, 
$\beta^a$ is a linear combination of $t^a$ and $x^a$; and furthermore
that, at $p_0$, 
$\beta^ax_a$ has positive derivative along $\gamma$, while
$\beta^at_a$ has zero derivative but negative second derivative.]
Note that the nonvanishing of $A^a$ at $p_0$ was used here,
to achieve $f_+ > 0$ and $f_- > 0$.
Finally, fix smooth spacelike slices, 
$S_-$, $S_0$, and $S_+$, passing through $p_-$, $p_0$, and $p_+$, 
respectively. 

Next, let there be given a neighborhood $U$ of $\gamma$ in $M$ and 
neighborhood $\hat{U}$ of $g_{ab}$ in $C^1[U]$, such that, with
respect to any metric $\tilde{g}_{ab}$ in $\hat{U}$, the vector field
$t^a$ continues to be timelike in $U$, and the surfaces $S_-, S_0$, 
and $S_+$ continue to be spacelike in $U$.   We write
``$\Theta \leadsto 0$" to mean ``given any $U$ and $\hat{U}$ as
above, the number $\Theta$ is bounded; and that bound can be made as 
small as we wish by choosing $U$ and $\hat{U}$ to be sufficiently small". 

Now let $U$ and $\hat{U}$ be given as above, and let $\tilde{g}_{ab}
\in \hat{U}$ satisfy conditions i)-iii) of the Theorem.  For $S$ 
any $\tilde{g}$-spacelike slice cutting $U$, and
$\xi^a$ any vector field in $U$, set
\begin{equation}
P(\xi, S) = \int_S\ \tilde{G}_{ab}\xi^b \tilde{dS}^a,
\label{int}\end{equation}
where $\tilde{G}_{ab}$ is the Einstein tensor, and $\tilde{dS}^a$ the 
surface-element, with respect to $\tilde{g}_{ab}$.  Set 
$m = P(t, S_0)$.  Then, by conditions i) and ii) of the
Theorem, $m > 0$.  We have, for $S$ and $S'$ any two 
$\tilde{g}$-spacelike slices cutting $U$, 
\begin{equation}
P(\xi, S) - P(\xi,S') = \int_V \tilde{G}_{ab}
(\tilde{g}^{c(a}\tilde{\nabla}_c\xi^{b)}) \tilde{dV},
\label{Sdiff}\end{equation}
where the integral is over the portion of $U$ between $S$ and $S'$,
and where we have used condition iii) of the Theorem.

This function $P(\ ,\ )$ has three properties of interest.

1.  $|P(t,S) - m|/m \leadsto 0$.  This follows from Eqn.
(\ref{Sdiff}).  Choose $S' = S_o$ and $\xi^a = t^a$ therein,
and use the energy condition and the fact that
$g^{c(a}\nabla_ct^{b)} = 0$ on $\gamma$.

2.  If $\xi^a$ vanishes at the point $S\cap\gamma$,
then $|P(\xi,S)|/m \leadsto 0$.  This follows from Eqn. (\ref{int}).
By choosing $U$ to be small, we may bound 
the components of $\xi^a$ on the right; 
while the energy condition implies that the integral of the components of 
$\tilde{G}_{ab}$ over $S$ is bounded by a multiple of $P(t,S)$,
and so, using property 1, by a multiple of $m$.

3.  If $g^{c(a}\nabla_c\xi^{b)}$  vanishes on $\gamma$, then 
$|P(\xi,S) - P(\xi,S')|/m \leadsto 0$.  This follows from
Eqn. (\ref{Sdiff}).  By choosing $U$ and $\hat{U}$ to be small, we may
bound the components of 
$\tilde{g}^{c(a}\tilde{\nabla}_c\xi^{b)}$ on the right; 
while the energy condition and property 1 imply that the integral of the
components of $\tilde{G}_{ab}$ over $V$ is bounded by a multiple
of $m$.

Now consider the following number
\begin{eqnarray}
K & = & P(\beta,S_+) + P(\beta, S_-) - 2 P(\beta, S_0) \\
& - & f_+ P(t, S_+) - f_- P(t, S_-) - g P(x, S_+) + g P(x, S_-). \nonumber
\label{1line}\end{eqnarray}
We estimate this number $K$ in two ways.  For the first, we use
property 2 above.  Combine the first, fourth, and sixth terms, 
using (\ref{alph+}) and this property; then combine the second, fifth, and 
seventh, using (\ref{alph-}) and this property; 
and finally apply to the third term  this property.
We conclude that $|K|/m \leadsto 0$.
For the second way, first combine the first three terms,
using property 3; then combine the last two, again using this
property; and finally apply to the two middle terms property 1. 
We conclude that $|K+(f_++  f_-)m|/m \leadsto 0$.
But, since the numbers $f_+$ and $f_-$ are positive, these two 
estimates contradict each other, completing the proof.$\ \ \Box$
\\

We remark that this theorem continues to hold if we weaken the hypothesis 
to require merely the existence, for each
suitable $\tilde{g}_{ab}$, of some $\tilde{g}$-conserved symmetric 
tensor field, $\tilde{T}_{ab}$, satisfying conditions i)-iii) of 
the Theorem.  This follows, since there was never used in the proof that 
the $\tilde{G}_{ab}$ in Eqn. (\ref{int}) is the Einstein tensor 
of $\tilde{g}_{ab}$.  Thus, for example, the theorem above is also
applicable to any metric theory of gravity.  

There is a version of the present theorem for Newtonian gravitation.
The curve $\gamma$ is replaced by a fixed time-parameterized curve in
space, the background metric $g_{ab}$ by a fixed Newtonian potential $\phi$,
the $\tilde{g}_{ab}$ by potentials $\tilde{\phi}$, the
$\tilde{G}_{ab}$ of Eqn. (\ref{int}) and its conservation by suitable 
matter fields and equations, and the conclusion by the
assertion that the acceleration
of this curve, at each of its points, is given by the value
of $-\nabla\phi$ 
there.  Just as it is not necessary (as described in the previous
paragraph) to impose Einstein's equation for the present theorem
in general relativity,
so it is not necessary to impose Poisson's equation for its 
Newtonian version.  This Newtonian result bounds the self-acceleration
of a body by the order of the ``acceleration of gravity" produced
by that body at its surface, e.g., for the Earth, by the order of 
10 m/sec$^2$.  A bound of the same order of magnitude is also
available in general relativity, as one sees by following through
the proof of the present theorem, keeping track of the inequalities. 

In Newtonian gravitation, there is available a much stronger result.  (See,
e.g., \cite{TD}, Eqn. (36).)  Denote by $\tilde{\phi}$ the actual
potential in which a body finds itself, and set $\phi$ equal to 
$\tilde{\phi}$ minus the self-potential of that body.  Then
the difference between the actual acceleration experienced
by the center of mass of that body 
and the ``free-fall acceleration" of the center of mass, 
$- \nabla\phi|_{cm}$, 
is bounded by the product of the body-size and the
variation in the external tidal field (i.e.,
in $\nabla\nabla\phi$) over that body.  For the Earth in its
orbit about the Sun,
that bound is less than $10^{-12}$ m/sec$^2$.  Unfortunately, 
it appears to be difficult to 
find a version of this Newtonian result in general relativity.  The key
problem, apparently, is that
there is in this theory no natural notion of ``center
of mass" or of ``self-potential" \cite{JE} \cite{TD} \cite{LB}.   

There may be a version of the present theorem applicable to charged
particles, but if there is one it will be somewhat more complicated.
For example, the given curve $\gamma$ and the ``background" electromagnetic
field together prescribe what is to be the limiting charge-to-mass 
ratio of the body; and so the hypothesis would have to be modified 
to require the existence only of bodies having that prescribed ratio.
The present theorem is not, of course, applicable to 
the motion of singular regions or of black holes, for there is no
natural ``background" with respect to which to describe the behavior
of such objects.


\begin{thebibliography} {99}

\bibitem{AE} A. Einstein and M. Grossmann, Ztschr.
Math. Phys. 62, 225 (1913).

\bibitem{JE} $\textit{Isolated gravitating systems in
General Relativity}$, J. Ehlers (ed.), Amsterdam 1979.

\bibitem{TD} Th. Damour in $\textit{300 years of gravitation}$, 
S.W. Hawking and W. Israel (eds.), CUP, 1987.

\bibitem{LB} L. Blanchet in LNP 540, B.G. Schmidt (ed.), Berlin
2000.

\bibitem{GJ} R.Geroch and P.S. Jang, $\textit{Motion of a body in 
General Relativity}$, J. Math. Phys. 16, 65-67 (1975).


\end{thebibliography}
\end{document}